\documentstyle[12pt,fleqn]{article}
\input epsf
\def\mpl{{{m_{Pl}}}}
\def\oph{{{\Omega_{\widetilde{\gamma}}h^2}}}
\def\be{\begin{equation}}
\def\ee{\end{equation}}
\def\ba{\begin{eqnarray}}
\def\ea{\end{eqnarray}}
\def\la{\mathrel{\mathpalette\fun <}}
\def\ga{\mathrel{\mathpalette\fun >}}
\def\fun#1#2{\lower3.6pt\vbox{\baselineskip0pt\lineskip.9pt
        \ialign{$\mathsurround=0pt#1\hfill##\hfil$\crcr#2\crcr\sim\crcr}}}
\def\pho{{{\widetilde{\gamma}}}}
\def\r0{{{R^0}}}
\def\glu{{{\widetilde{g}}}}

\def\GeV{{{\mbox{GeV}}}}
\def\MeV{{{\mbox{MeV}}}}
\def\SUSY{{{{\sc susy}}}}
\def\LSP{{{{\sc lsp}}}}
\def\LEP{{{{\sc lep}}}}
\def\LROCS{{{{\sc lrocs}}}}
\def\WIMPS{{{{\sc wimps}}}}
\def\cm{{{\mbox{cm}}}}
\def\photino{{{\mbox{photino}}}}
\def\gluino{{{\mbox{gluino}}}}
\def\mb{{{\mbox{mb}}}}
\def\avg#1{{{{\langle #1 \rangle }}}}

\def\mpl{{{m_{Pl}}}}
\def\re#1{{[\ref{#1}]}}
\def\eqr#1{{Eq.\ (\ref{#1})}}
\def\mst{{{M_{\widetilde{S}}}}}
\begin{document}
\begin{titlepage}
\null\vspace{-72pt}
\begin{flushright}
{\footnotesize
FERMILAB--Pub--95/068-A\\
RU-95-18\\
astro-ph/9504081\\
April 1995 \\
Submitted to {\em Phys.\ Rev.\ D}}
\end{flushright}
\renewcommand{\thefootnote}{\fnsymbol{footnote}}
\vspace{0.15in}
\baselineskip=24pt

\begin{center}
{\Large \bf  Light photinos as dark matter}\\
\baselineskip=14pt
\vspace{0.75cm}
Glennys R.\ Farrar\footnote{Electronic mail: {\tt farrar@farrar.rutgers.edu}}\\
{\em Department of Physics and Astronomy\\
Rutgers University, Piscataway, NJ~~08855}\\
\vspace{0.4cm}
Edward W.\ Kolb\footnote{Electronic mail: {\tt rocky@rigoletto.fnal.gov}}\\
{\em NASA/Fermilab Astrophysics Center\\
Fermi National Accelerator Laboratory, Batavia, IL~~60510, and\\
Department of Astronomy and Astrophysics, Enrico Fermi Institute\\
The University of Chicago, Chicago, IL~~ 60637}\\
\end{center}

\baselineskip=24pt

\begin{quote}
\hspace*{2em}
There are good reasons to consider models of low-energy supersymmetry
with very light photinos and gluinos.  In a wide class of models the
lightest $R$-odd, color-singlet state containing a gluino, the $\r0$,
has a mass in the 1-2 GeV range and the slightly lighter photino,
$\pho$, would survive as the relic $R$-odd species.  For the light
photino masses considered here, previous calculations resulted in an
unacceptable photino relic abundance.  But we point out that processes
other than photino self-annihilation determine the relic abundance
when the photino and $R^0$ are close in mass.  Including
$\r0\longleftrightarrow\pho$ processes, we find that the photino relic
abundance is most sensitive to the $\r0$-to-$\pho$ mass ratio, and
within model uncertainties, a critical density in photinos may be
obtained for an $\r0$-to-$\pho$ mass ratio in the range 1.2 to 2.2. We
propose photinos in the mass range of 500 MeV to 1.6 GeV as a dark
matter candidate, and discuss a strategy to test the hypothesis.

\vspace*{8pt}

PACS number(s): 98.80.Cq, 14.80.Ly, 11.30.Pb

\renewcommand{\thefootnote}{\arabic{footnote}}
\addtocounter{footnote}{-2}
\end{quote}
\end{titlepage}

\newpage

\baselineskip=24pt
\renewcommand{\baselinestretch}{1.5}
\footnotesep=14pt

\vspace{36pt}
\centerline{\bf I. INTRODUCTION}
\vspace{24pt}

In this paper we study the early-Universe evolution and freeze out of
light, long-lived or stable, $R$-odd states, the photino $\pho$ and
the gluino $\glu$.\footnote{$R$-parity is a multiplicative quantum
number, exactly conserved in most \SUSY\  models, under which
ordinary particles have $R=+1$ while new ``superpartners'' have
$R=-1$. Throughout this paper we will assume that $R$-parity is exact
so the lightest $R$-odd particle is stable.}   In the type of models
we consider, the photino should be the relic $R$-odd particle, even
though it may be more massive than the gluino.  This is because below
the confinement transition the gluino is bound into a color-singlet
hadron, the $R^0$, whose mass (which is in the 1 to 2 GeV range when
the gluino is very light [\ref{EXPTS},\ref{PHENO}]) is greater than
that of the photino.  Including previously neglected reactions
associated with the gluino (more precisely, associated with the
$R^0$), we find that light photinos may be cosmologically acceptable;
indeed they are an attractive dark-matter candidate.

In the minimal \SUSY\ model, the mass matrix of the charged and neutral
 \SUSY\  fermions (gauginos and Higgsinos) are determined by Lagrangian
terms involving the Higgs chiral superfields, $\widetilde{H}_1$ and
$\widetilde{H}_2$, and the SU$(2)$ and U$(1)$ gauge superfields,
$\widetilde{W}^a$ and $\widetilde{B}$, plus soft supersymmetry
breaking terms.  This leads to a neutralino mass matrix in the basis
$(\widetilde{B},\   \widetilde{W}^3,\ \widetilde{H}_1^0,\
\widetilde{H}_2^0)$ of the form
\be
\label{eq:MM} \hspace*{-12pt}
\left( \!  \begin{array}{cccc}
M_1 & 0       & -m_Z\cos\beta\sin\theta_W  &  m_Z \sin\beta\sin\theta_W    \\
0       & M_2 &  m_Z\cos\beta\cos\theta_W & -m_Z \sin\beta\cos\theta_W  \\
-m_Z\cos\beta\sin\theta_W &   m_Z \cos\beta\cos\theta_W & 0       & -\mu  \\
 m_Z\sin\beta\sin\theta_W   & -m_Z \sin\beta\cos\theta_W   & -\mu &   0
\end{array} \! \right) \!  .
\ee
Here $m_Z$ is the mass of the $Z$-boson, $\theta_W$ is the Weinberg
angle, $\mu$ is the coefficient of a supersymmetric mixing term
between the Higgs superfields, and $\tan\beta$ is the ratio of the
vacuum expectation values of the two Higgs fields responsible for
electroweak symmetry breaking.  The \SUSY-breaking masses $M_1$ and
$M_2$ are commonly assumed to be of order $m_Z$ or larger, and if the
\SUSY\ model is embedded in a grand unified theory, then $3M_1/M_2 =
5\alpha_1/\alpha_2$.

The terms in the Lagrangian proportional to $M_1$ and $M_2$ arise from
dimension-3 \SUSY-breaking operators.  However such \SUSY-breaking
terms are not without problems.  It appears difficult to break \SUSY\
dynamically in a way that produces dimension-3 terms while avoiding
problems associated with the addition of gauge-singlet superfields
\re{BKN}.  In models where \SUSY\ is broken dynamically or
spontaneously in the hidden sector and there are no gauge singlets,
all dimension-3 \SUSY-breaking operators in the effective low-energy
theory are suppressed by a factor of $\avg{\Phi}/\mpl$, where
$\avg{\Phi}$ is the vacuum expectation value of some hidden-sector
field.  Thus, dimension-3 terms effectively do not contribute to the
neutralino mass matrix.  This would imply that at the tree level the
gluino is massless, and the neutralino mass matrix is given by
\eqr{eq:MM} with vanishing $(00)$ and $(11)$ entries.  However, a
non-zero gluino mass, as well as non-zero entries in the neutralino
mass matrix are generated through radiative corrections such as the
top-stop loop, and for the neutralinos, ``electroweak'' loops
involving higgsinos and/or winos and binos.

The generation of radiative gaugino masses in the absence of
dimension-3 \SUSY\  breaking was studied by Farrar and Masiero
\re{FM}.\footnote{See also \re{pierce_papa} for general formulae.
Earlier studies [\ref{bgm},\ref{bm}] of radiative corrections when
tree level gaugino masses are absent included another dimension-3
operator, the so-called ``$A$ term,'' and did not consider the
electroweak loop contributions to the neutralino mass matrix.  They
also assumed model-dependent relations between parameters.} From
Figs.\ 4 and 5 of that paper one sees that as $M_0$, the typical
\SUSY-breaking scalar mass, varies between 100 and 400 GeV, the gluino
mass ranges from about $700 $ to about 100 MeV,\footnote{Actually,
larger values of $M_0$ are not considered in order to keep the gluino
mass greater than about 100 MeV.  Otherwise an unacceptably light
pseudoscalar meson would be produced \re{EXPTS}.} while the
photino\footnote{Upon diagonalization of the mass matrix, the physical
neutralino states are a linear combination of $\widetilde{B}^0$,
$\widetilde{W}^3$, $\widetilde{H}_1^0$, and $\widetilde{H}_2^0$.  When
the gaugino submatrix elements are small, the lightest neutralino is a
linear combination of $\widetilde{W}^3$ and $\widetilde{B}^0$ that is
almost identical to the SU$(2)$ $\times$ U$(1)$ composition of the
photon, and thus is correctly called ``photino.''}  mass ranges from
around 400 to 900 MeV, for $\mu \ga 40$ GeV.  This estimate for the
photino mass should be considered as merely indicative of its possible
value, since an approximation for the electroweak loop used in Ref.\
\re{FM} is strictly valid only when $\mu$ or $M_0$ are much larger
than $m_W$.  The other neutralinos are much heavier, and the
production rates of the photino and the next-lightest neutralino in
$Z^0$ decay are consistent with \LEP\ bounds \re{FM}.

Using the results of Ref.\ \re{FM}, but additionally restricting
parameters so that the correct electroweak symmetry breaking is
obtained, Farrar \re{PHENO} found $M_0 \sim 150 $ GeV and estimated
the $R^0$ lifetime.  This allowed completion of the study of the main
phenomenological features of this scenario, which was begun in Ref.\
\re{EXPTS}.  The conclusion is that light gluinos and
photinos are quite consistent with present experiments, and result in
a number of striking predictions \re{PHENO}.  However models with
light gauginos have been widely thought to be disallowed because it has
been believed that the relic density of the lightest neutralino,
usually referred to as the \LSP,\footnote{In this scenario, \LSP\ is an
ambiguous term: the gluino is lighter than the photino, although the
photino is lighter than the $R^0$.  A more relevant term would be
\LROCS---{\em lightest $R$-odd color singlet}.} exceeds cosmological
bounds unless $R$-parity is violated [\ref{G}--\ref{LSP}].

In this paper we point out that previous considerations of the relic
abundance have neglected the rather important interplay between the
photino and the gluino which can determine the final neutralino
abundance if the photino and gluino are both light, as they must be in
models without dimension-3 explicit \SUSY-breaking terms.   We find that
when gluino--photino interactions are included, rather than being a
cosmological embarrassment, these
very light photinos are an excellent dark matter candidate.  In this
paper we discuss the decoupling and relic abundance of light photinos,
and the sensitivity of the result upon the parameters of the \SUSY\
models.

For the light masses studied here, freeze-out occurs well after the
confinement transition so the physical states must be color singlets.
Since $\glu$ is not a color singlet, below the confinement transition
the relevant state to consider is the lightest color-singlet state
containing a gluino, which is believed to be a gluon-gluino bound
state known as the $\r0$.  The other light $R$-odd states are more
massive than these, and decay to the two light ones with lifetimes
much faster than the expansion rate at freeze out.  The only other
possible state of interest is the $S^0$, which is the lightest $R$-odd
baryon, consisting of the color-singlet, flavor-singlet state
$uds\glu$ [\ref{EXPTS},\ref{PHENO},\ref{GF:84}].  The masses for the
states we consider will be assumed to be in the range
[\ref{PHENO},\ref{FM}]
\begin{eqnarray}
\glu  (\gluino):       & &  m_\glu = 100 - 600  \ \MeV     \nonumber \\
\pho (\photino):    & &  m = 100 \ \MeV  - 1400  \ \MeV      \nonumber \\
\r0(\glu g):             & &  M = 1 - 2                     \ \GeV
\nonumber \\
S^0(uds\glu):        & &  M_{S^0} = 1.5 - 2     \ \GeV.
\end{eqnarray}
Since it is the lightest {\it color-singlet} $R$-odd state, the $\pho$
is stable, and $\r0$ decays to a final state consisting of a photino
and typically one meson: $\r0 \longrightarrow \pho\pi;~\pho\eta$, etc.
The lifetime is very uncertain, but probably lies in the range
$10^{-4}$ to $10^{-10}$s, or even longer \re{PHENO}.

The reaction rates that determine freeze out will depend upon the
$\pho$ and $\r0$ masses, the cross sections involving the $\pho$ and
$\r0$, and possibly the decay width of the $\r0$ as well.  In turn
the cross sections and decay width also depend on the masses of the
$\pho$, $\glu$ and $\r0$, as well as the masses of the squarks and
sleptons.  We will denote the squark/slepton masses by a common mass
scale $\mst$ (expected to be of order 100 GeV).  Even if the masses
were known and the short-distance physics specified, calculation of
the width and some of the cross sections would be no easy task,
because one is dealing with light hadrons.  Fortunately, our
conclusions are reasonably insensitive to individual masses,
lifetimes, and cross sections, but depend crucially upon the
$\r0$-to-$\pho$ mass ratio.

When we do need an explicit value of the photino mass $m$ or the
masses of squarks and sleptons, we will parameterize them by the
dimensionless ratios
\be
\label{eq:UNCERTAIN}
\mu_8  \equiv  \frac{m}{800\ \MeV}; \qquad
\mu_{S} \equiv \frac{\mst}{100\ \GeV}\ .
\ee

Although there are several undetermined parameters in our calculation,
as mentioned above, the most important parameter will be the the ratio
of the $\r0$ mass to the $\pho$ mass:
\be
r\equiv \frac{M}{m}.
\ee
This is by far the most crucial parameter, with the relic abundance
having an exponential dependence upon $r$.  We find that limits to the
magnitude of the contribution to the present mass density from relic
photinos requires $r \la 2.2,$\footnote{Or else $R$-parity must be
violated so the photinos decay.} while $r$ must be larger than about
1.2 if the photino relic density is to be significant.  This narrow
band of $r$ encompasses the large uncertainties in lifetimes and cross
sections.  If the mass ratio is between about 1.6 and 2, then
light-mass relic photinos could dominate the Universe and provide the
dark matter with $\Omega_\pho \sim 1$.

In the concluding section we explore the proposal that light photinos
are the dark matter, and discuss possibilities for testing the idea.
We lay the groundwork for this suggestion in the next section as we
develop a new scenario for decoupling and freeze-out for the photinos
and gluinos.  In Section III we consider the cross sections and
lifetimes used in Section IV to calculate the reaction rates relevant
for the determination of the
freeze-out abundance of the photinos (and hence $\Omega_\pho h^2$).
In Section V we compare the reaction rates to the expansion rate and
estimate when photinos decouple.

\newpage
\centerline{\bf II. SCENARIO FOR PHOTINO/GLUINO FREEZE OUT}
\vspace{24pt}

The standard procedure for the calculation of the present number
density of a thermal relic of the early-Universe is to assume that the
particle species was once in thermal equilibrium until at some point
the rates for self-annihilation and
pair-creation processes became much smaller than the expansion rate,
and the particle species effectively froze out of equilibrium.  After
freeze out, its number density decreased only because of the dilution
due to the expansion of the Universe.   (For a discussion, see Ref.\  \re{KT}.)

Since subsequent to freeze out the number of particles in a {\em co-moving}
volume is constant,  it is convenient to express the number density of
the particle species in terms of the entropy density, since the
entropy in a co-moving volume is also constant for most of the history
of the Universe.  The number density-to-entropy ratio is usually
denoted by $Y$.  If a species of mass $m$ is in equilibrium and
non-relativistic, $Y$ is simply given in terms of the
mass-to-temperature ratio $x\equiv m/T$ as
\be
\label{eq:EQAB}
Y_{EQ}(x) = 0.145 (g/g_*)x^{3/2}\exp(-x),
\ee
where $g$ is the number of spin degrees of freedom, and $g_*$ is the
total number of relativistic degrees of freedom in the Universe at
temperature $T=m/x$.  Well after freeze-out $Y(x)$ is constant, and we
will denote this asymptotic value of $Y$ as $Y_\infty$.

If self annihilation determines the final abundance of a species,
$Y_\infty$ can be found by integrating the Boltzmann equation (dot
denotes $d/dt$)
\be
\dot{n} + 3Hn = -\avg{|v|\sigma_A}\left(n^2-n_{EQ}^2\right),
\ee
where $n$ is the actual number density, $n_{EQ}$ is the equilibrium density,
$H$ is the expansion rate of the Universe, and $\avg{|v|\sigma_A}$ is the
thermal average [\ref{KR},\ref{GG}] of the annihilation rate.

There are no general closed-form solutions to the Boltzmann equation,
but there are reliable, well tested approximations for the late-time
solution, i.e., $Y_\infty$.  Then with knowledge of $Y_\infty$, the
contribution to $\Omega h^2$ from the species can easily be found.
Let us specialize to the survival of photinos assuming
self-annihilation determines freeze out.

Calculation of the relic abundance involves first calculating the
value of $x$, known as $x_f$, where the abundance starts to depart
from the equilibrium abundance.  Using standard approximate solutions
to the Boltzmann equation \re{KT} gives\footnote{Freeze-out aficionados
will notice that we use the formulae appropriate for $p$-wave
annihilation because Fermi statistics requires the initial identical
Majorana fermions to be in an $L=1$ state \re{G}.}
\be
\label{eq:XFO}
x_f =    \ln(0.0481\mpl m \sigma_0) -
1.5 \ln[\ln(0.0481\mpl m \sigma_0)] \ ,
\ee
where we have used $g=2$ and $g_*=10$, and parameterized the
non-relativistic annihilation cross section as $\avg{|v|\sigma_A} =
\sigma_0 x^{-1}$.  In anticipation of the results of the next section,
we use $\sigma_0=2\times10^{-11} \mu_8^2\mu_S^{-4}\mb$, and we find $
x_f \simeq 12.3 + \ln(\mu_8^3/\mu_S^4)$.  The value of $x_f$
determines $Y_\infty$:
\be
\label{eq:YINFTY}
Y_\infty = \frac{2.4 x_f^2}{\mpl m \sigma_0} \simeq
7.4\times10^{-7}\mu_8^{-3}\mu_S^4.
\ee

Once $Y_\infty$ is known, the present photino energy density can be
easily calculated: $ \rho_\pho= m n_\pho = 0.8\, \mu_8\GeV \cdot
Y_\infty\, 2970\, \cm^{-3}$.  When this result is divided by the critical
density, $\rho_C = 1.054h^2\times 10^{-5}\ \GeV\, \cm^{-3}$, 
the fraction of the critical density contributed by the photino is $\oph =
2.25\times10^8 \mu_8 Y_\infty$.  For $Y_\infty$ in \eqr{eq:YINFTY},
$\oph = 167\mu_8^{-2}\mu_S^4$.

The age of the Universe restricts $\oph$ to be less than one, so for
$\mu_S=1$, the photino must be more massive than  about  10 GeV if its relic
abundance is determined by self-annihilation.

But in this paper we point out that for models in which both
the photino and the gluino are light, freeze-out is not determined by
photino self annihilation, but by $\pho$--$\r0$ interconversion.  The
basic point is that since the $\r0$ has strong interactions, it will
stay in equilibrium longer than the photino, even though it is more
massive. As long as $\pho \leftrightarrow \r0$ interconversion occurs
at a rate larger than $H$, then through its interactions with the
$\r0$ the photino will be able to maintain its equilibrium abundance
even after self-annihilation has frozen out.\footnote{Actually,
interconversion can also play an important role in determining the relic
abundance of heavier photinos.  When the photino is more massive and
freeze-out occurs above the confinement phase transition, the
analysis is similar to the one here;  in fact simpler because perturbation
theory can be used to compute the relevant rates involving gluinos and
photinos.  Since the qualitative relation between interconversion and
self-annihilation rates is independent of whether the gluino is free
or confined in an $R^0$, one can get a crude idea of the required
gluino-photino mass ratio, $r$,  just by using the analysis in this
paper and scaling the results to the value of $\mu_8$ of interest.
We concentrate on the light gaugino scenario because it is attractive
in its own right, and also because it {\it naturally} produces $r$ in
the right ballpark\re{PHENO}.  In a conventional \SUSY-breaking scheme
fine-tuning is generally necessary to give $r$ the right value for the
interconversion mechanism to play an important role.}

Before we demonstrate that this scenario naturally occurs for the
types of photino and $\r0$ masses expected, we must determine the
cross sections and decay width of the reactions involving the photino
and the gluino.

\vspace{36pt}
\centerline{\bf III. CROSS SECTIONS AND DECAY WIDTH}
\vspace{24pt}

In this section we characterize the cross sections and decay width
required for the determination of the relic photino abundance, and
also discuss the uncertainties.  We should emphasize that all cross
sections are calculated in the non-relativistic (N.R.) limit, and by
$\avg{\cdots}$ we imply that the quantity is to be evaluated as a
thermal average [\ref{KR},\ref{GG}].   In the
N.R.\ limit  a temperature dependence to the cross sections
enters if the annihilation proceeds through a $p$-wave, as required if
the initial state consists of identical fermions \re{G}.  For $p$-wave
annihilation, at low energy the cross section is proportional to
$v^2$, where $|v|$ is the relative velocity of the initial particles.
The thermal average reduces to replacing $v^2$ by $6T/m$, where $m$ is
the mass of the particle in the initial state.

We now consider in turn the cross sections and width for the individual
reactions discussed in the previous section.

\vspace{30pt}

\centerline{\bf A.\ \ Self-annihilations and co-annihilation}

The first type of reactions we will consider are those which change
the number of $R$-odd particles.

\underline{$\r0\r0\rightarrow X$}: We will refer to this process as $\r0$
self-annihilation.  At the constituent level the relevant reactions are
$\glu + \glu \rightarrow g + g$ and $\glu + \glu \rightarrow q +
\bar{q}$, which are unsuppressed by any powers of $\mst$, and should be
typical of strong interaction cross sections.  In the N.R.\ limit, we
expect the $\r0\r0$ annihilation cross section to be
comparable to the $\bar{p}p$ cross section, but with an extra factor
of $v^2$, accounting for the fact that there are identical fermions in
the initial state, so annihilation must proceed through a
$p$-wave.\footnote{In general the result is not so simple.  For instance, in
addition to the term proportional to $v^2$,  the
cross section also involves a term proportional to the square of the
masses of the initial and final particles. }
There is some energy dependence to the $\bar{p}p$ cross section, but
it is sufficient to consider $\avg{|v|\sigma_{\r0\r0}}$ to be a
constant, approximately given by
\be
\avg{|v|\sigma_{\r0\r0}} \simeq 100v^2 \, \mb = 600\,  x^{-1}\,  r^{-1}\, \mb ,
\ee
where we have used for the relative velocity $v^2=6T/M=6/(rx)$, with
$ x\equiv m/T$.

We should note that the thermal average of the cross section might be
even larger if there are resonances near threshold.  In any case, this
cross section should be much larger than any cross section involving
the photino, and will ensure that the $\r0$ remains in equilibrium
longer than the $\pho$, greatly simplifying our considerations.

\underline{$\pho\pho \rightarrow X$}:  In photino self-annihilation at
low energies the final state $X$ is a lepton-antilepton pair, or
a quark-antiquark pair which appears as light mesons.  The process
involves the $t$-channel exchange of a virtual squark or slepton
between the photinos, producing the final-state fermion-antifermion
pair.  In the low-energy limit the mass $\mst$ of the squark/slepton
is much greater than $\sqrt{s}$, and the
photino-photino-fermion-antifermion operator appears in the low-energy
theory with a coefficient proportional to $e_i^2/\mst^2$, with $e_i$
the charge of the final-state fermion.\footnote{The electric charge
$e$ and the strong charge $g_S$ are to be evaluated at a scale of
order $\mst$, so in numerical estimates we use $\alpha_{EM}=1/128$ and
$\alpha_S=0.117$.} Also, as there are two identical fermions in the
initial state, the annihilation proceeds as a $p$-wave, which
introduces a factor of $v^2$ in the low-energy cross section \re{G}.
The resultant low-energy photino self-annihilation cross section is
[\ref{G},\ref{LSP},\ref{INDIRECTA},\ref{KKC}]:
\be
\label{eq:pho-self}
\avg{ |v|\sigma_{\pho\pho} }  =
8\pi\alpha_{EM}^2\,  \sum_iq_i^4\, \frac{m^2}{\mst^4}\,  \frac{v^2}{3}
\simeq 2.0\times10^{-11}\  x^{-1}   \ \left[\mu_8^2 \mu_{S}^{-4}\right] \ \mb,
\ee
where we have used for the relative velocity $v^2=6/x$ with $ x\equiv
m/T$, and $q_i$ is the magnitude of the charge of a final-state
fermion in units of the electron charge.  For the light photinos we
consider, summing over $e$, $\mu$, and three colors of $u$, $d$, and
$s$ quarks leads to $\sum_i q_i^4= 8/3$.

\underline{$\pho\r0 \rightarrow X$}: This is an example of a phenomenon
known as co-annihilation, whereby the particle of interest (in our case the
photino) disappears by annihilating with another particle (here, the $\r0$).
Of course co-annihilation also leads to a net decrease in $R$-odd particles.

In all processes involving the photino--$\r0$ interaction, the leading
tree-level short-distance operator containing $\glu$ and $\pho$ is
$\lambda^{\dagger}_{\glu} \lambda_{\pho}q_i^{\dagger}q_i + h.c.$, with
coefficient $e q_i g_S/\mst^2$.  For three light quarks,
$\sum_i q_i^2=2$.  Thus we can estimate the cross section for
$\pho\r0 \rightarrow X$ in terms of the $\pho$ self annihilation cross
section:
\be
\avg{ |v|\sigma_{\pho\r0} } \simeq \frac{\alpha_S}{\alpha_{EM}} \frac{4}{3}
\frac{2}{8/3} \frac{M}{m} \frac{3}{v^2}\ \avg{ |v|\sigma_{\pho\pho} } ,
\ee
where the ratio of $\alpha$'s arises because the short-distance operator
for co-annihilation is proportional to $e_i^2 g_S^2$ rather than
$e_i^4$, the second factor is the color factor coming from the gluino
coupling,
and the third factor comes from the ratio of $\sum_i q_i^2/\sum_i q_i^4$
for the participating fermions.  We have replaced $m^2$ appearing in
\eqr{eq:pho-self} by $mM$, although the actual dependence on $m$ and
$M$ may be more complicated.  Finally, the annihilation is $s$-wave so
there is no $v^2/3$ suppression as in photino self-annihilation.

Although the short-distance physics is perturbative, the initial
gluino appears in a light hadron, and there are complications in the
momentum fraction of the $\r0$ carried by the gluino and other
non-perturbative effects.  For our purposes it will be sufficient
to account for the uncertainty by including in the cross
section an unknown coefficient $A$, leading to a final expression
\be
\avg{ |v|\sigma_{\pho\r0} }  \simeq  1.5\times10^{-10} \ r\
                \left[\mu_8^2 \mu_{S}^{-4}A\right] \ \mb .
\ee
It is reassuring that if one estimates $\avg{ |v|\sigma_{\pho\r0} }$
starting from $\avg{ |v|\sigma_{\r0\r0} }$ a similar result is
obtained.   We find that co-annihilation will not be important unless
$A$ is larger than $10^2$ or so, which we believe is unlikely.

\vspace{30pt}

\centerline{\bf B.\ \ $\pho$--$\r0$ interconversion}

In what we call interconversion processes, there is an $R$-odd particle
in the initial as well as the final state.  Although the reactions do not
of themselves change the number of  $R$-odd particles, they keep the photinos
in equilibrium with the $\r0$s, which in turn are kept in
equilibrium through their self annihilations.

\underline{$\r0 \rightarrow \pho \pi$}:  $\r0$ decay can occur via,
e.g., the gluino inside the initial $\r0$ turning into an antiquark
and a virtual squark, followed by squark decay into a photino and a
quark.  In the low-energy limit the quark--antiquark--gluino--photino
vertex can be described by the same type of four-Fermi interaction as
in co-annihilation.  One expects on dimensional grounds a decay width
$\Gamma_0 \propto \alpha_{EM}\alpha_S M^5/ \mst^4$.  The lifetime of a
free gluino to decay to a photino and massless quark-antiquark pair
was computed in Ref.\ \re{HK}.  However this does not provide a very
useful estimate when the gluino mass is less than the photino mass.

The lifetime for $\r0$ decay was studied in Ref.\ \re{PHENO}.  In an
attempt to account for the effects of gluino-gluon interactions in
the $\r0$, necessary for even a crude estimate of the $R^0$
lifetime, the following picture was developed:  The $\r0$ is viewed as
a state with a massless gluon carrying momentum fraction $x$, and a
gluino carrying momentum fraction $(1-x),$\footnote{Of course there
should be no confusion with the fact that in the discussion of the
$\r0$ lifetime we use $x$ to denote the gluon momentum fraction
whereas throughout the rest of the paper $x$ denotes $m/T$.}  having
therefore an effective mass $M \sqrt{1-x}$.  The gluon structure
function $F(x)$ gives the probability in an interval $x$ to $x + dx$
of finding a gluon, and the corresponding effective mass for the
gluino.  One then obtains the $R^0$ decay width (neglecting the mass
of final state hadrons):
\be
\Gamma(M,r) = \Gamma_0(M,0) \int^{1-r^{-2}}_0 \! dx \
(1-x)^{5/2} F(x) \  f(1/r\sqrt{1-x}),
\ee
where $\Gamma_0(M,0)$ is the rate for a gluino of mass $M$ to decay to
a massless photino, and $f(y)=[(1-y^2)(1 + 2y - 7y^2 + 20y^3 - 7 y^4 +
2 y^5 + y^6) + 24y^3(1 - y + y^2)\log y]$ contains the phase space
suppression which is important when the photino becomes massive in
comparison to the gluino.  Modeling $K^{\pm}$ decay in a similar
manner underestimates the lifetime by a factor of 2 to 4.  This is in
surprisingly good agreement; however caution should be exercised when
extending the model to $R^0$ decay, because kaon decay is much less
sensitive to the phase-space suppression from the final state masses
than the present case, since the range of interest will turn out to be
$r\sim 1.2-2.2$.  For $r$ in this range, taking $F(x) \sim 6 x(1-x)$
following the discussion in Ref.\ \re{PHENO} leads to an approximate
behavior
\be
\label{eq:FR}
\Gamma_{\r0\rightarrow\pho\pi} =
            2.0\times10^{-14} \ {\cal F}(r)\ \GeV\  [\mu_8^5\mu_S^{-4}B] \ ,
\ee
where ${\cal F}(r)=r^5(1 - r^{-1})^6$, and the factor $B$ reflects the
overall uncertainty.  We believe a reasonable range for $B$ is
$1/30 \la B \la 3$.

\underline{$\r0\pi \longleftrightarrow \pho\pi$}:  We will refer to these
processes as photino-$R^0$ conversion, since an initial
$R^0$ (or $\pho$) is converted to a final  $\pho$ (or $\r0$).  The
short-distance
subprocess in this reaction is $q + \glu \rightarrow q + \pho$, again
described by the same low-energy effective operator as in
co-annihilation and $R^0$ decay.  At the hadronic level the matrix
element for $R^0 \pi \rightarrow \pho X$ is the same as for $\r0 \pho
\rightarrow \pi X$ for any $X$, evaluated in different physical
regions.  Thus the difference between the various cross sections is
just due to the difference in fluxes and final state phase space
integrations, and variations of the matrix element with kinematic
variables.  Given the crude nature of the analysis here, and the great
uncertainty in the overall magnitude of the cross sections,
incorporating the constraints of crossing symmetry are not justified
at present.

We can point to one specific hadronic effect which we do not include
but which is potentially important.  It is likely that near threshold
there is a resonance (the $R_{\pi}$) which would increase the cross
section by a factor of $4 M_R^2/\Gamma_R^2$, where $M_R$ is the mass
and $\Gamma_R$ the width of the resonance.  This complicates matters
because neither the resonance's width nor its distance above threshold
is known.  If a resonance is important, it would also be necessary to
perform the thermal average over the resonance in a more careful
manner \re{GG}.

We will parameterize our uncertainty by including a factor $C$ in the
cross section to express the uncertainty due to hadronic physics and
the possible existence of a resonance.  Putting everything together,
we obtain
\be
\label{eq:EQA}
\avg{ |v|\sigma_{\r0\pi} } \simeq  1.5\times10^{-10}\, r  \,
\left[\mu_8^2 \mu_{S}^{-4}C\right] \ \mb  .
\ee
We would expect $C$ to fall in the range $1 \la C \la 10^{3}$.  We will use
detailed balance arguments which allow us to avoid using the inverse
reaction, $\pho \pi \rightarrow \r0 \pi$.

This completes the discussion of the lifetimes, cross sections,  and their
uncertainties.  The results are summarized in Table I.

\begin{table}
\footnotesize{\hspace*{1em} Table I: Cross sections and the decay width
used in the calculation of the relic photino abundance.  The
dimensionless parameters $\mu_8$ and $\mu_S$ were defined in Eq.\
(\ref{eq:UNCERTAIN}), and ${\cal F}(r)$ was discussed below
\eqr{eq:FR}.  The coefficients $A$, $B$, and $C$ reflect uncertainties
involving the calculation of hadronic matrix elements. }
\begin{center}
\begin{tabular}{ll|l} \hline\hline
\multicolumn{2}{c|}{Process}  & Cross section or width \\ \hline
$\r0$ self annihilation: & $\avg{|v|\sigma_{\r0\r0}} $ &
$ 600\  x^{-1}\  r^{-1}\ \mb$ \\
$\pho$ self annihilation: & $\avg{ |v|\sigma_{\pho\pho} } $ &
$ 2.0\times10^{-11}\  x^{-1}\ \left[\mu_8^2 \mu_{S}^{-4}\right]\   \mb$\\
co-annihilation: & $\avg{ |v|\sigma_{\pho\r0} }   $ &
$1.5\times10^{-10}\  r  \        \left[\mu_8^2 \mu_{S}^{-4}A\right]\ \mb$ \\
$\r0$ decay: & $\Gamma_{\r0\rightarrow\pho\pi} $ &
$ 2.0\times10^{-14}\  {\cal F}(r)\ [\mu_8^5\mu_S^{-4} B] \ \GeV $ \\
$\pho$ -- $\r0$ conversion: & $\avg{ |v|\sigma_{\r0\pi} } $ &
$ 1.5 \times10^{-10}\  r \ \left[\mu_8^2 \mu_{S}^{-4}C\right]\ \mb$ \\
\hline \hline \end{tabular}
\end{center}
\end{table}

\vspace{36pt}
\centerline{\bf IV. EARLY-UNIVERSE REACTION RATES}
\vspace{24pt}

To obtain an estimate of when the rates will drop below the expansion
rate, we will assume all particles are in LTE (local thermodynamic
equilibrium).  In LTE a particle of mass $m$ in the N.R.\ limit has a
number density
\be
\label{eq:EQ}
n = \frac{g}{(2\pi)^{3/2}}(mT)^{3/2}\exp(-m/T)
=\frac{g}{(2\pi)^{3/2}}(T/m)^{3/2}m^3\exp(-m/T) .
\ee
Here $g$ counts the number of spin degrees of freedom, and will be $2$
for the $\r0$ and the $\pho$.

\underline{ $H$ (the expansion rate)}: Of course all rates are to be
compared with the expansion rate.  In the radiation-dominated Universe
with $g_*\sim 10$ degrees of freedom
\be
H  =  1.66 g_*^{1/2} T^2/\mpl =  2.8 \times 10^{-19}x^{-2}  \left[
\mu_8^2\right] \   \GeV.
\ee

\underline{$\pho\pho\rightarrow X$ (photino self-annihilation)}:  In the
Boltzmann equation for the evolution of the $\pho$ number density
there are terms accounting for photino self-annihilation and photino
pair-production from light particles in the plasma.  Assuming the
light annihilation products are in LTE, the terms are of the form
\be
\dot{n}_\pho + 3Hn_\pho \supset - \avg{|v|\sigma_{\pho\pho}}\left[
\left(  n_\pho\right)^2  - \left(n_\pho^{EQ}\right)^2 \right]   \  .
\ee
If we assume that the photino is in equilibrium, the self-annihilation
and pair production terms are equal, and we may express the individual
terms in the form
\be
\dot{n}_\pho \supset  -3Hn_\pho \mp \left[\avg{|v|\sigma_{\pho\pho}}
        n_\pho^{EQ} \right] n_\pho^{EQ} \ ,
\ee
where the upper sign is for self-annihilation and the lower sign is for
pair production.

It is obvious that $\left[n_\pho^{EQ}
\avg{|v|\sigma_{\pho\pho}}\right]$ plays the role of a ``rate'' to be
compared to $H$.  If this rate is much greater than $H$, the
self-annihilation/pair-production processes will ensure the photino is
in equilibrium, while if the rate is much smaller than $H$,
self-annihilation/pair-production cannot enforce equilibrium.

Therefore we define an equilibrium photino annihilation rate by
$\Gamma(\pho\pho\rightarrow X) = n_\pho^{EQ}\avg{|v|\sigma_{\pho\pho}}$.
Using \eqr{eq:EQ} for the equilibrium abundance and the annihilation
cross section discussed in the previous section, we find
\begin{eqnarray}
\Gamma(\pho\pho\rightarrow X) & = &  \frac{2}{(2\pi)^{3/2}}
\left(\frac{T}{m}\right)^{3/2}  m^3 \exp(-m/T)
   \  \frac{2.0\times10^{-11} \mb}{0.39\ \mb \, \GeV^2} \ x^{-1} \
                \left[\mu_8^2 \mu_{S}^{-4}\right] \nonumber \\
     & = &  3.3\times10^{-12} x^{-5/2} \exp(-x)
                    \left[\mu_8^5 \mu_{S}^{-4}\right] \ \GeV .
\end{eqnarray}

\underline{$\r0\r0\rightarrow X$ ($\r0$ self-annihilation)}:
Determination of the equilibrium rate for $\r0$-self-annihilation proceeds
in a similar manner, yielding $\Gamma(\r0\r0\rightarrow X) =n_\r0^{EQ}
\avg{|v|\sigma_{\r0\r0}}$:
\begin{eqnarray}
\Gamma(\r0\r0\rightarrow X)  & = &  \frac{2}{(2\pi)^{3/2}}
\left(\frac{T}{M}\right)^{3/2} M^3
          \exp(-M/T) \  \frac{240\ x^{-1}\, r^{-1}\, \mb}{0.4\ \mb \, \GeV^2}
\nonumber \\
        & = & 99\,  r^{1/2} x^{-5/2} \exp(-r x) \  [\mu_8^3] \ \GeV .
\end{eqnarray}

\underline{$\pho\r0 \rightarrow X$ ($\pho$ co-annihilation)}:  In the
Boltzmann equation for the evolution of the $\pho$ density will appear
a term $-n_\r0 n_\pho  \avg{|v|\sigma_{\pho\r0}}$.  Therefore the equilibrium
co-annihilation rate for the decrease of the $\pho$ density is
\begin{eqnarray}
\Gamma(\pho\r0\rightarrow X) & = &  n_\r0^{EQ}  \  \avg{|v|\sigma_{\pho\r0}}
                                                                \nonumber \\
        & = &   2.5\times10^{-11}\ r^{5/2}\ x^{-3/2}\ \exp(-rx) \
[\mu_8^5\mu_S^{-4}]\   ,
\end{eqnarray}
where we have again assumed the particles in the process are in equilibrium.

\underline{$\pho \pi \rightarrow \r0$ (Inverse decay)}:  If the $\r0$
decay products (in this case $\pho$ and $\pi$) are in equilibrium,
then the Boltzmann equation for the evolution of $\r0$ contains a term
\be
\dot{n}_\r0 + 3Hn_\r0 \supset -\Gamma_{\r0\rightarrow\pho\pi}
        \left( n_\r0 - n^{EQ}_\r0 \right)  \  .
\ee
The first term on the rhs represents decay, while the second term
represents ``inverse decay.''  Since inverse decay turns a $\pho$ into
a $\r0$, there will be an ``inverse decay'' term in the equation for
the evolution of the $\pho$ number density:
\be
\dot{n}_\pho + 3Hn_\pho \supset - \Gamma_{\r0\rightarrow\pho\pi}\ n^{EQ}_\r0 \
{}.
\ee
The right hand side can be written as $ n^{EQ}_\pho (n^{EQ}_\r0/n^{EQ}_\pho)\
\Gamma_{\r0\rightarrow\pho\pi}$.  Therefore the inverse decay
rate in the evolution of the photino number density contributes a term
\begin{eqnarray}
\Gamma(\pho\pi\rightarrow \r0) & = &  \Gamma_{\r0\rightarrow\pho\pi}\
       (n^{EQ}_\r0/n^{EQ}_\pho) \nonumber  =  \Gamma_{\r0\rightarrow\pho\pi}
       \left( \frac{M}{m}\right)^{3/2}  \exp[-(M-m)/T]\nonumber \\
 & = & 2.0\times10^{-14}\ r^{3/2} \ {\cal F}(r) \ \exp\left[-(r-1)x\right] \
                        [\mu_8^5\mu_S^{-4}B] \ \GeV .
\end{eqnarray}

\underline{$\pho\pi \rightarrow \r0\pi$ (photino--$R^0$ conversion)}:
It is easiest to obtain this term by first considering the term in the
equation for $\dot{n}_\r0$ due to the reverse process and then using
detailed balance:
\be
\dot{n}_\r0 \supset - n_\pi n_\r0\avg{|v|\sigma_{\r0\pi}}   \ .
\ee
Since the photino-$R^0$ conversion process creates a $\pho$ there is a
similar term in $\dot{n}_\pho$ with the opposite sign.  Now we can
write this in a form to calculate the rate for $\pho$ annihilation by
\be
\dot{n}_\pho\supset -\dot{n}_\r0 = n_\pi n_\r0\avg{|v|\sigma_{\r0\pi}}
            = \left[  \frac{n_\pi n_\r0}{n_\pho}\,
                     \avg{|v|\sigma_{\r0\pi}}\right]n_\pho  \ .
\ee

Assuming equilibrium as before, the rate keeping the $\pho$ in equilibrium can
be expressed as
\begin{eqnarray}
\Gamma(\pho\pi\rightarrow \r0\pi) & = & \frac{n_\r0^{EQ}}{n_\pho^{EQ}}
                n_\pi^{EQ}  \avg{|v|\sigma_{\r0\pi}} \nonumber \\
& = & 2.7\times10^{-12}\ r^{5/2}\ x^{-3/2}\ \exp(-0.175\mu_8^{-1}x) \nonumber
\\
& & \times  \exp[-(r-1)x]\ [\mu_8^{7/2}\mu_S^{-4}C]  \ .
\end{eqnarray}

Of course it is the ratio of the reaction rates to the expansion rate
that will be used to estimate photino freeze out.  These  ratios are given
in Table II.

\begin{table}
\footnotesize{\hspace*{1em} Table II: The ratio of the equilibrium
reaction rates to the expansion rate for the indicated reactions.
Shown in $[\cdots]$ is the scaling of the rates with unknown
parameters characterizing the cross sections and decay width.}
\begin{center}
\begin{tabular}{ll|ll} \hline \hline
\multicolumn{2}{c|}{Process}  & $\Gamma/H$ & \\ \hline
$\pho$ self annihilation & $\pho\pho\rightarrow X$
& $1.2\times10^7\  x^{-1/2}\,\exp(-x)$
                       & $[\mu_8^3  \mu_S^{-4}]$ \\
$\r0$ self annihilation & $\r0\r0\rightarrow X$
& $ 3.5\times10^{20} x^{-1/2}\, r^{1/2}\, \exp(-rx) $
                       & $[\mu_8]$ \\
co-annihilation & $\pho\r0\rightarrow X $
& $8.9\times10^7\  x^{1/2}\, r^{5/2}\,\exp(-rx)$
                       & $[\mu_8^3 \mu_S^{-4}A]$ \\
inverse decay & $\pho\pi\rightarrow\r0$
& $7.1\times10^4\  x^2\, r^{3/2}\, {\cal F}(r)\ \exp[-(r-1)x]$
                       & $[\mu_8^{3}\mu_S^{-4}B]$ \\
$\pho$ -- $\r0$ conversion & $\pho\pi\rightarrow\r0\pi$
& $ 9.6\times10^6\  x^{1/2}\, r^{5/2}\, \exp[-(r-1)x]\,
                                        \exp(-0.175\mu_8^{-1}x)$
                       &  $ [\mu_8^{3/2}\mu_S^{-4}C] $ \\
\hline \hline \end{tabular}
\end{center}
\end{table}

There are two striking features apparent when comparing the magnitudes
of the equilibrium reaction rates in Table II.  The first feature is
that the numerical factor in $\r0$ self annihilation is enormous in
comparison to the other numerical factors.  This simply reflects the
fact that $\r0$ annihilation proceeds through a strong process, while
the other processes are all suppressed by a factor of $\mst^{-4}$.

The other important feature is the exponential factors of the rates.
They will largely determine when the photino will decouple, so it is
worthwhile to examine them in detail.

The exponential factor in $\pho$ self-annihilation is simply $e^{-m/T}$,
which arises from the equilibrium abundance of the $\pho$.  It is simple to
understand: the probability of one $\pho$ to find another $\pho$ with
which to annihilate is proportional to the photino density, which
contains a factor of $e^{-m/T}$ in the N.R.\ limit.

The similar exponential factor in $\r0$ self-annihilation is also easy to
understand. An $\r0$ must find another $\r0$ to annihilate, and that
probability is proportional to $e^{-M/T}=e^{-rx}$.

Co-annihilation is also an exothermic process, so the only exponential
suppression is the probability of a $\pho$ locating the $\r0$ for
co-annihilation, proportional to the equilibrium number density of
$\r0$, in turn proportional to $e^{-M/T}=e^{-rx}$

In inverse-decay the exponential factor is $e^{-(r-1)x}=e^{-(M-m)/T}$.
The number density of target pions is $e^{-m_\pi/T}$, so this factor is
present.  It is
necessary for the $\pi$--$\pho$ collision to have sufficient
center-of-mass energy to create the $\r0$.  This introduces an addition
suppression of $e^{-(M-m-m_\pi)/T}$.  Combining the two exponential
factors gives the result in Table II.

\begin{figure}[p] \vspace*{-24pt}
\hspace*{25pt}  \epsfxsize=400pt \epsfbox{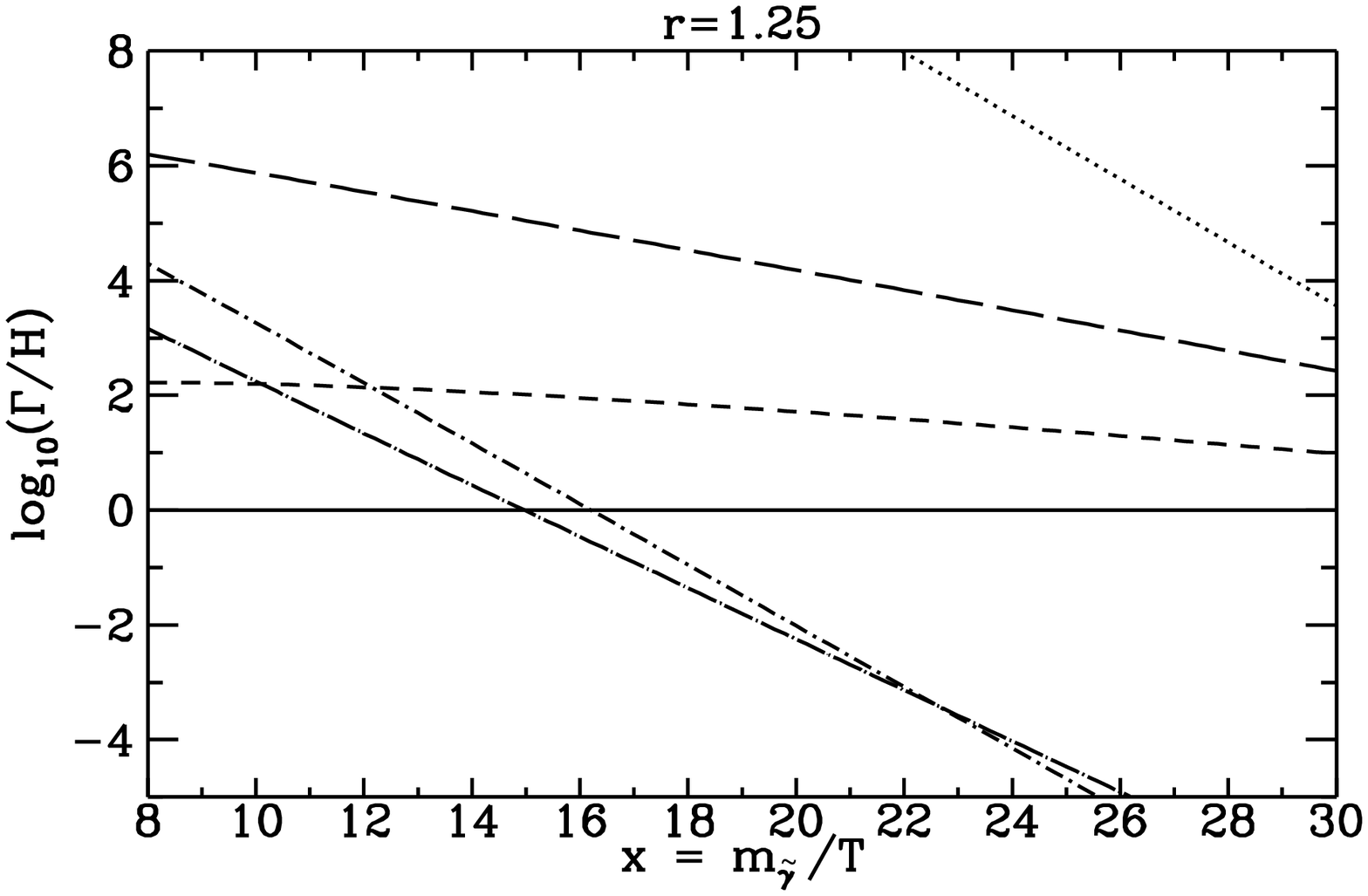}\\
\hspace*{25pt}\epsfxsize=400pt \epsfbox{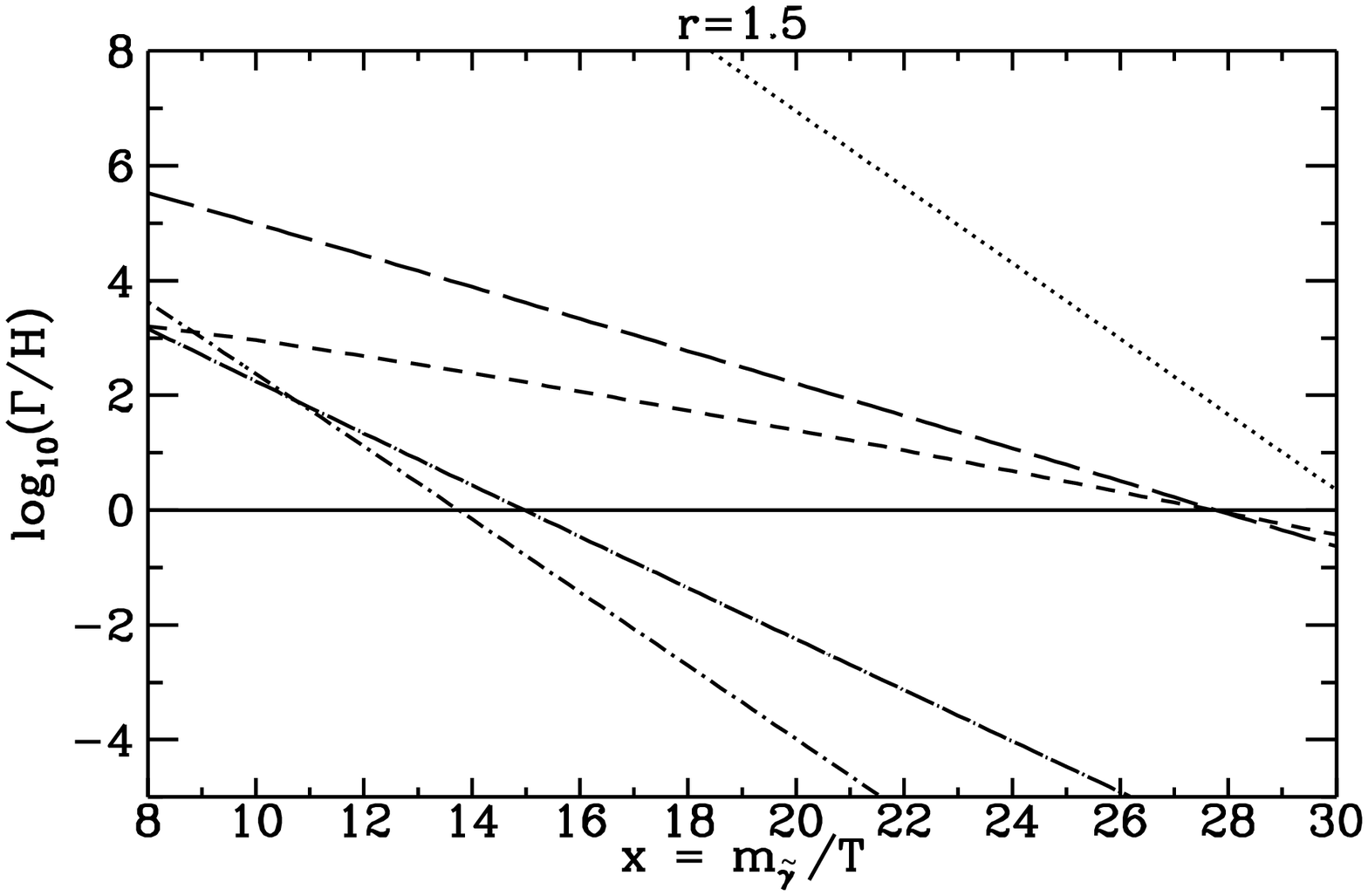}\\
\hspace*{45pt} \epsfxsize=360pt \epsfbox{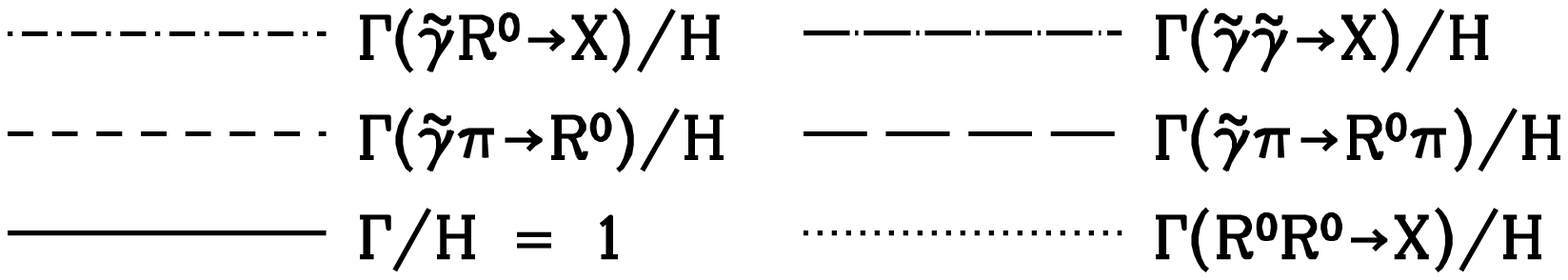} \vspace*{-12pt} \\
\hspace*{1em}\footnotesize{{\bf Fig.\ 1:}  Equilibrium reaction rates divided
by $H$ for $r=1.25$ and 1.5, assuming $\mu_8=\mu_S=1$, and that the factors
$A=B=C=1$.  The rates can be easily scaled for other choices of the
parameters. }
\end{figure}

\begin{figure}[p] \vspace*{-24pt}
\hspace*{25pt}\epsfxsize=400pt \epsfbox{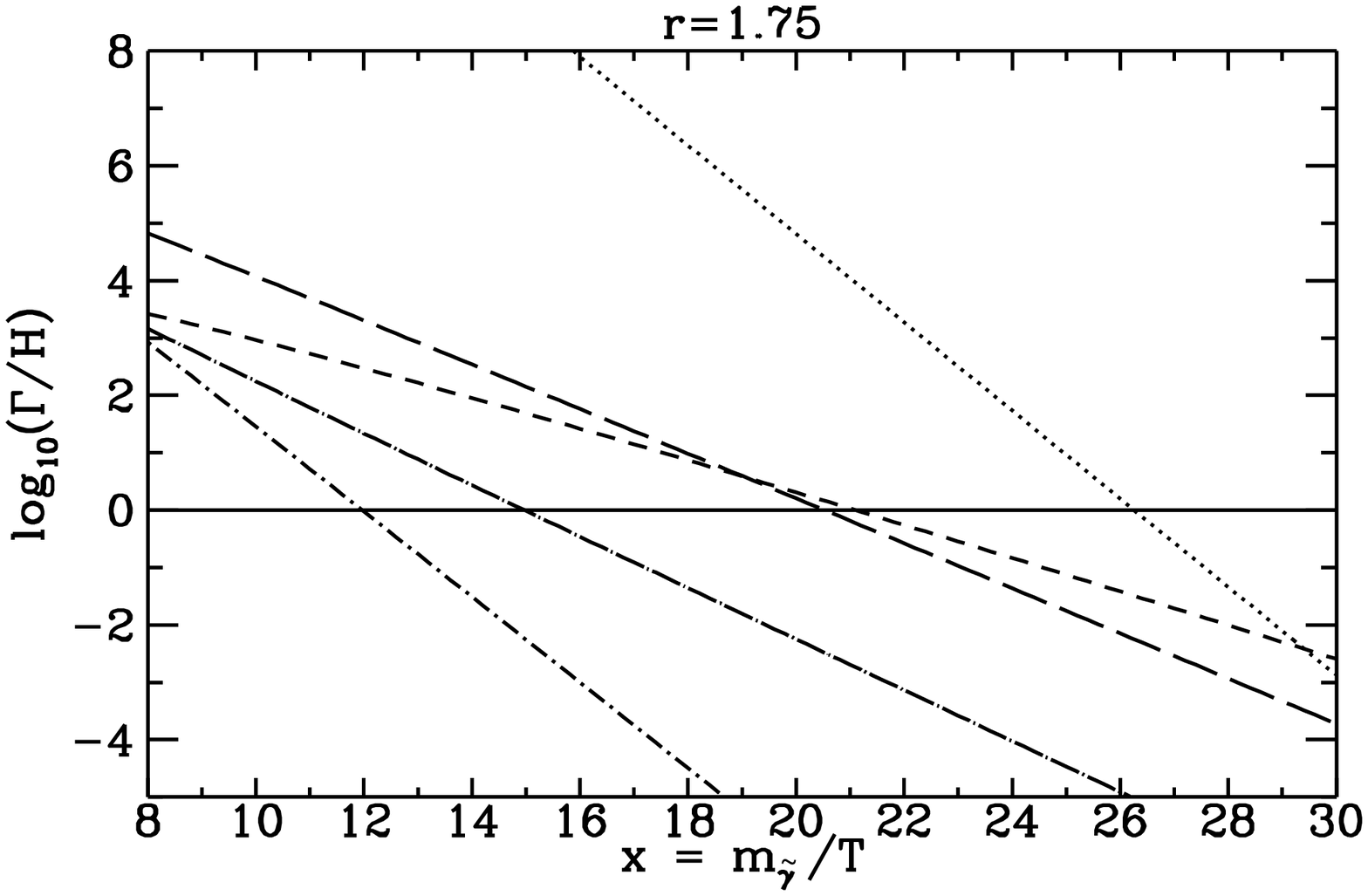}\\
\hspace*{25pt}\epsfxsize=400pt \epsfbox{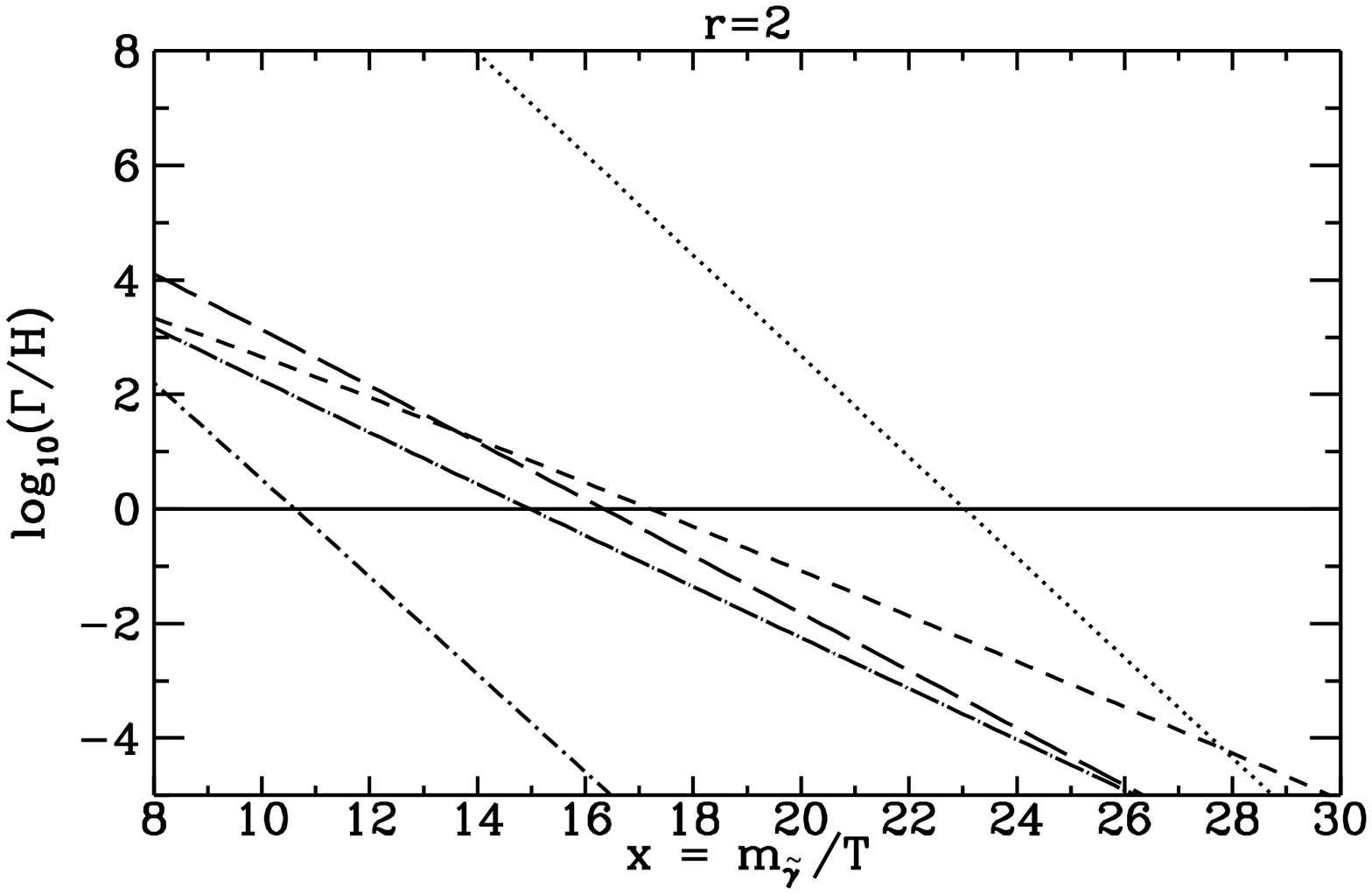} \\
\hspace*{45pt} \epsfxsize=360pt \epsfbox{legend.eps} \vspace*{-12pt} \\
\hspace*{1em}\footnotesize{{\bf Fig.\ 2:}  The same as Fig.\ 1, but for
$r=1.75$ and 2. }
\end{figure}

Finally, photino-gluino conversion involves two exponential
suppression factors.  The first, $e^{-m_\pi/T} =e^{-0.175\mu_8^{-1}x}$
represents the suppression in the pion number density,\footnote{At the
temperatures of interest for decoupling, pions might be cheap, but they are
not free.} and since the mass of the $\r0$ is greater than the mass of
the $\pho$, there is an additional $e^{-(M-m)/T}$ suppression.

The factors of $x$ and $r$ originate from three places: a factor of
$x^2$ comes from dividing the rates by $H$, factors of $r$ and $x$ arise
for pre-exponential factors in the number density, and finally they may
appear explicitly in the cross section or decay width.

The equilibrium reaction rates divided by $H$ are shown in Figs.\ 1
and 2 for $r=1.25$, 1.5, 1.75, and 2.  In the figures we have assumed
$\mu_8=\mu_S=A=B=C=1$.  Using the information in Table II it is possible
to scale the curves for other values of the parameters.

\vspace{36pt}
\centerline{\bf V. ANALYSIS}
\vspace{24pt}

Rather than integrate a complete reaction network for the evolution
and freeze out of the photinos, we will assume that the photinos
remain in equilibrium so long as there is a reaction depleting the
$\pho$ abundance that is larger than $H$.  We will then assume that as
soon as the rate of the last such reaction drops below $H$, the
photinos immediately freeze out, and the photino-to-entropy ratio is
frozen at that value.  We will call this approximation the ``sudden''
approximation.

We can get some idea of the accuracy of the sudden approximation by
considering a simple system involving only photino self-annihilation.
As discussed in Section II, there is a well developed formalism for
calculating the self-annihilation freeze-out of a N.R.\ species
\re{KT}.  Using that formalism in Section II, \eqr{eq:YINFTY}, we
found $Y_\infty \simeq 7.4 \times10^{-7}$.

Now let us compute $Y_\infty$ using the sudden approximation.  From
Fig.\ 1 or Fig.\ 2, we see that $\Gamma(\pho\pho\rightarrow
X)=H$ at $x=14.7$, independent of $r$.  We will denote by $x_*$ the
value of $x$ when $\Gamma=H$.  Using the sudden approximation that the
$\pho$ is in LTE until $x=x_*$ and immediately freezes out would give
a photino to entropy ratio of (again using 2 degrees of freedom and
$g_*=10$)
\be
\label{eq:SUDDEN}
Y_\infty = Y_{EQ}(x_*) = 0.145\, (2/10) x_*^{3/2}\exp(-x_*) =7\times10^{-7}
\quad (\mbox{using}\  x_*=14.7).
\ee

The agreement between $Y_\infty$ obtained using the sudden
approximation, \eqr{eq:SUDDEN}, and the usual freeze-out calculation,
\eqr{eq:YINFTY}, suggests that the sudden approximation is a
reasonable point of departure for a first look at this phenomenon.
Note, however, that the accuracy of the sudden approximation when
self-annihilation is the principal photino equilibration mechanism
does not guarantee that it is an equally good approximation when
interconversion is the important process.  The Boltzmann equation when
photino self-annihilation dominates can be written \re{KT}
\be
\label{eq:boltzself}
\frac{dY}{dx} =  \frac{- x \avg{\sigma_{\pho \pho}|v|} s}{H(m)}  ( Y^2 -
Y_{EQ}^2 ),
\ee
where $Y_{EQ}(x)$ has the form given in Eq. \ref{eq:SUDDEN} and $H(m)
= 1.67 g_*^{1/2}m^2/m_{pl}$.  This is to be contrasted with the
analogous expression when interconversion dominates:
\be
\label{eq:boltzinter}
\frac{dY}{dx} =  \frac{- x \avg{\sigma_{\pho \pi \rightarrow R^0 \pi}|v|}
s}{H(m)}
( Y - Y_{EQ} ) Y_{\pi}.
\ee
Here, $Y_{\pi}$ is the equilibrium pion to entropy ratio:
\be
\label{eq:Ypi}
Y_{\pi}(x) = 0.145 (3/2) (2/10) (r_{\pi})x^{3/2}\exp(- r_{\pi}x).
\ee
We have introduced $r_{\pi} \equiv m_{\pi}/m=
0.175\mu_8^{-1}$, and included the factor $(3/2)$ because the pion has
three flavor$\times$spin degrees of freedom in comparison to the
photino's two.  The difference in these forms, in particular the much
weaker exponential dependence on $x$ for $Y_{\pi}$ compared to
$Y_{EQ}$, is largely responsible for the shallower slope of the
interconversion and inverse-decay curves as compared to the
self-annihilation curves in Figs.\ 1 and 2. This shallower slope
means that the quality of the sudden approximation in this
case is inferior to the self-annihilation case, but probably not
significantly in comparison to the large uncertainty due to our
present poor knowledge of the cross sections.  Closer examination of
this question is in progress \re{CFK}.

\begin{figure}
\hspace*{25pt} \epsfxsize=400pt \epsfbox{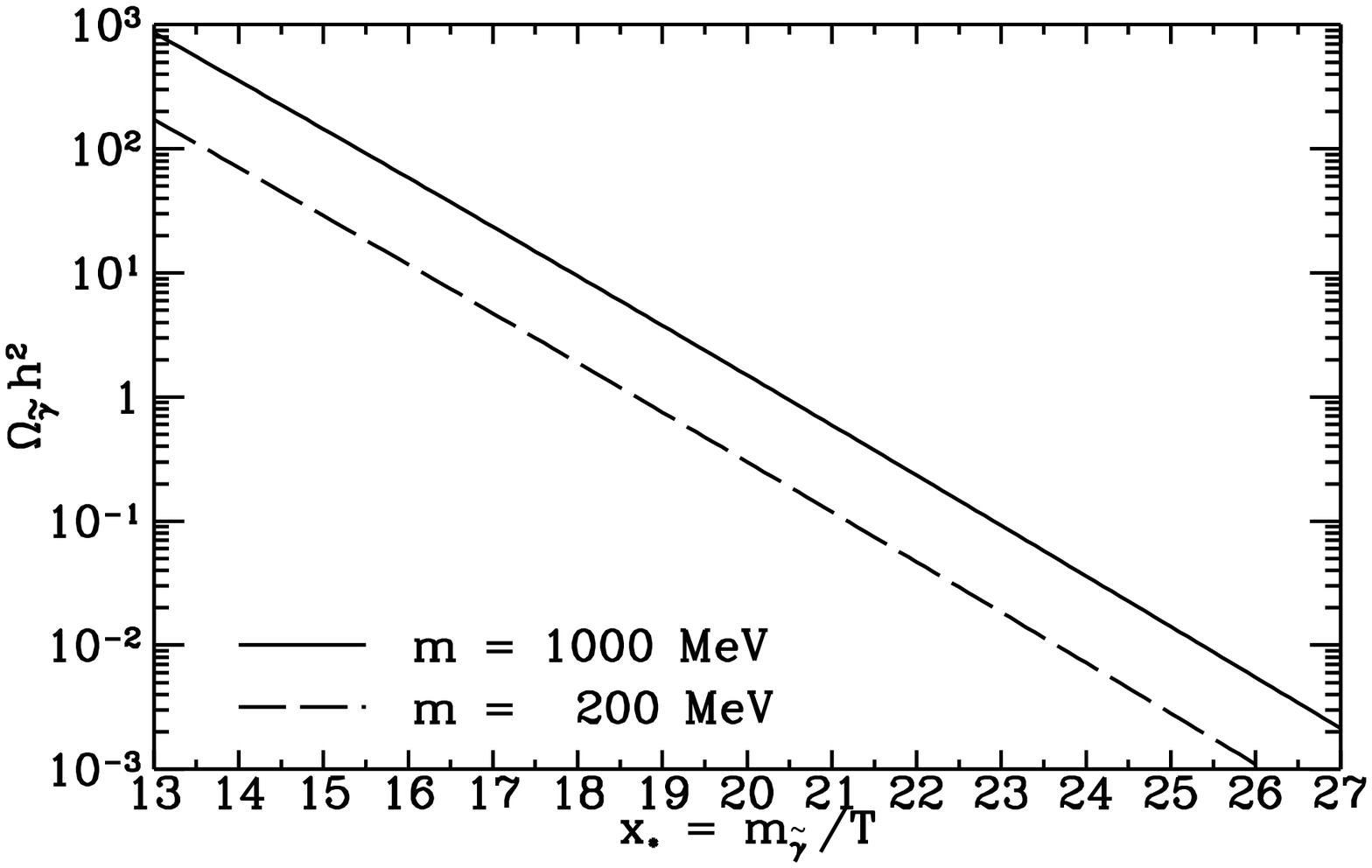}\\
\hspace*{1em} {\footnotesize{{\bf Fig.\ 3:} $\Omega_\pho h^2$ as a
function of $x_*$ assuming the photino stays in equilibrium until $x_*$ and
immediately decouples (the sudden approximation). } }
\end{figure}

Now we proceed using the sudden approximation.  Given $x_*$,
we wish to determine $\Omega_\pho h^2$.  It is, of course, a
very sensitive function of $x_*$:  $\Omega_\pho h^2 =2.25\times10^{8}\
[\mu_8]\ Y_\infty=6.5 \times10^6\ [\mu_8]\ x_*^{3/2}\exp(-x_*)$.  The
dependence of $\oph$ upon $x_*$ is shown in graphical form in Fig.\ 3,
with specific values presented in Table III.

\begin{table}[bt]
\footnotesize{\hspace*{1em} Table III:  The value of  $\Omega_\pho
h^2$ assuming sudden freeze out at $x=x_*$.   $\Omega_\pho h^2=1$ occurs
around $x_*=20$, and $\Omega_\pho h^2 = 10^{-2}$ around $x_*=25$.  In the Table
we have taken $\mu_8=1$. }
\begin{center}
\begin{tabular}{l|cccccccccccc} \hline\hline
$x_*$ &12 & 14 & 16 & 18 & 20 & 21 & 22 &  24 & 25 & 26 & 28 &30\\ \hline
$\Omega_\pho h^2$  & 1660 & 283 & 47 & 7.6  & 1.2  & 0.5 & 0.2  &
0.03
&  0.01 & 0.004 & 0.0007& $10^{-6}$ \\
\hline\hline \end{tabular}
\end{center}
\end{table}

Since the age of the universe restricts $\Omega_\pho h^2$ to be
smaller than unity, $x_*$ must be larger than 20.  In order for the
relic photinos to be dynamically interesting in structure evolution
$\oph$ must be larger than $10^{-2}$, which obtains for $x_*\la 25$.
Photinos would dominate the mass of the Universe if $\oph\ga
0.03$,\footnote{Nucleosynthesis bounds the contribution from baryons
to be about $\Omega_B h^2\la 0.03$.} which would result if $x_*=24$.
For $\Omega_\pho=1$ and $h\sim 1/2$, $x_*$ must be about 22. Thus we
can summarize interesting values of $x_*$ by
\be \begin{array}{lll}
\phantom{20 \la}\  x_* \la 25;   & \oph\ga10^{-2}; &\mbox{dynamically
                        interesting role for photinos}  \\
\phantom{20 \la}\  x_*\la 24;    &  \oph\ga 0.03; & \mbox{photinos dominate
baryons}  \\
20 \la x_* \la 23;\hspace{1em} &  \oph\sim 0.9,& \mbox{photinos are the dark
matter and}\ \Omega_{TOT=1}  \\
\phantom{20 \la}\  x_* \la 20;   &  \oph\ga1;& \mbox{disallowed by age
arguments}.
\end{array}
\ee

\begin{figure}
\hspace*{25pt} \epsfxsize=400pt \epsfbox{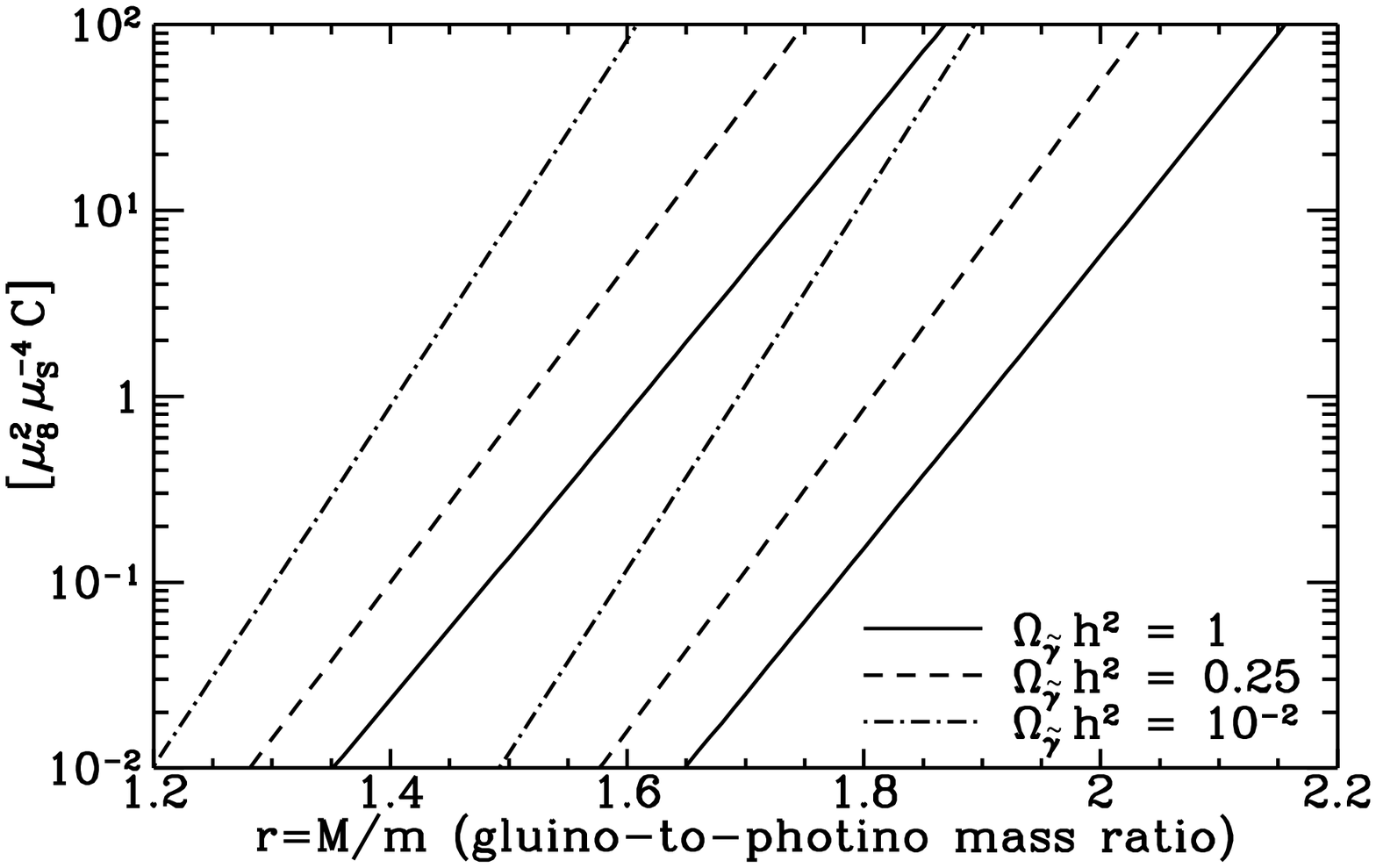}\\
\hspace*{1em} {\footnotesize{{\bf Fig.\ 4:} Assuming $\pho$ freeze out is
determined by $\pho$--$\r0$ conversion, the figure shows as a function of $r$
the values of $[\mu_8^2\mu_S^{-4}C]$ required to give the indicated values of
$\Omega_\pho h^2$.  The uncertainty band is generated by allowing $\mu_8$ to
vary
independently over the range $0.5 \leq \mu_8 \leq 2$. } }
\end{figure}

Now in turn, $x_*$ is exponentially sensitive to $r=M/m$, so limits to
the contribution to the present density from $\pho$ will be a sensitive
probe of $r$.

{}From Figs.\ 2 and 3, we see that for the canonical choices $\mu_8 =
\mu_S = A = B = C =1$, either the interconversion process, or
decay-inverse decay is the last photino reaction to be of importance.
It is impossible to say which one because of the uncertainties in the
computation of the cross section and the decay width, so we shall
consider both possibilities in turn.

If interconversion determines the relic abundance  and we make the
sudden approximation  then we can determine $\oph$ as a function of
the unknown parameters.  Such a graph is given in Fig.\ 4.
{}From the graph we see that $\oph<1$ can result for $r=2.2$ if we allow
$\mu_8^2\mu_S^{-4}C$ to be as large as $10^2$.  We also see that a
dynamically interesting value of $\oph$ can result for $r$ as small as
1.2 if $\mu_8^2\mu_S^{-4}C=10^{-2}$, although if the interconversion
rate is suppressed this much, it is likely inverse decay would govern
freeze out.

\begin{figure}
\hspace*{25pt} \epsfxsize=400pt \epsfbox{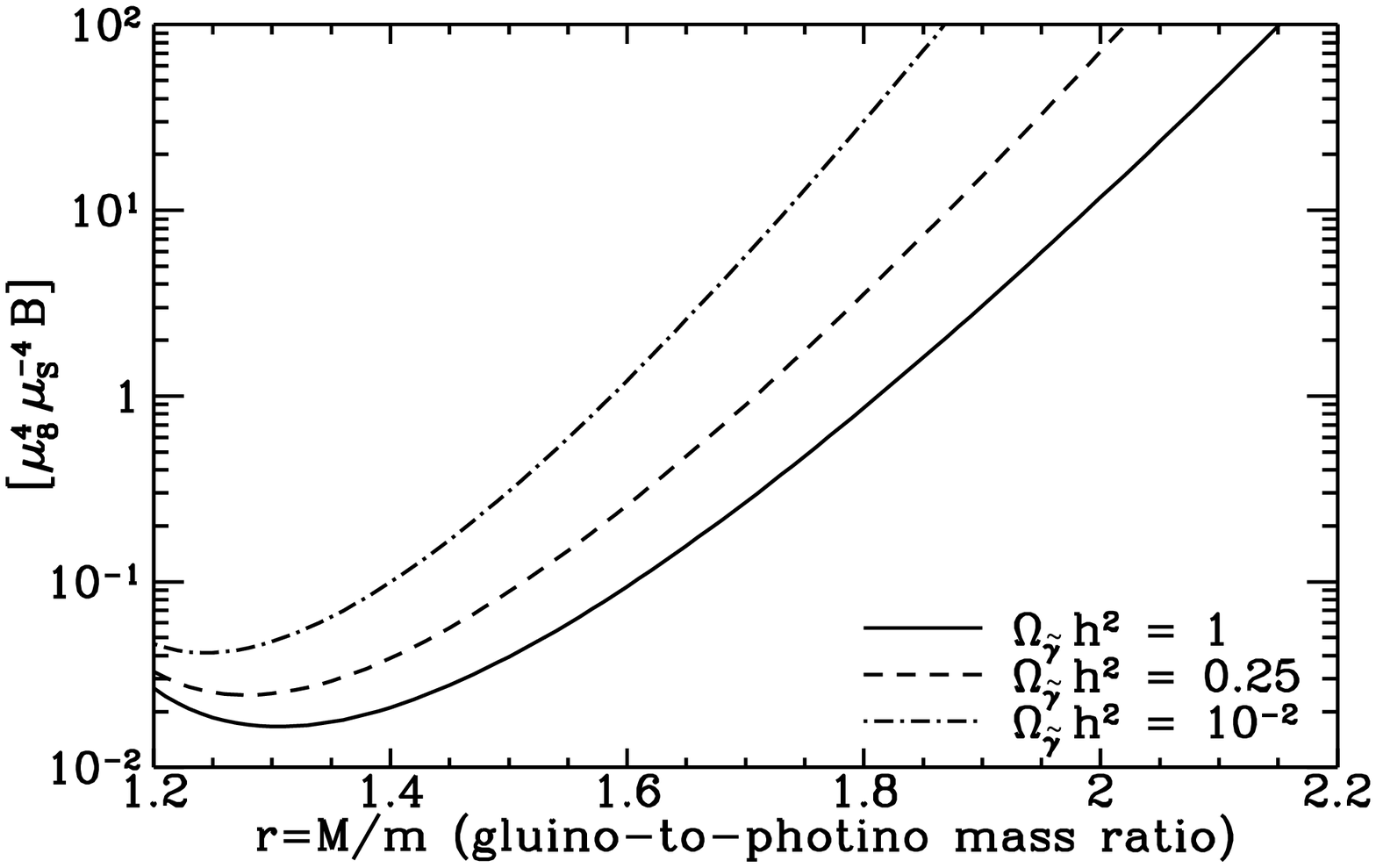}\\
\hspace*{1em} {\footnotesize{{\bf Fig.\ 5:} Assuming $\pho$ freeze out is
determined by decay/inverse decay, the figure shows as a function of $r$
the values of $[\mu_8^4\mu_S^{-4}B]$ required to give the indicated
values of $\Omega_\pho h^2$.   } }
\end{figure}

A similar calculation can be made assuming that inverse decay is the
last operative reaction depleting the photinos.  The result of such an
analysis is shown in Fig.\ 5.  For $r\ga1.4$ the behavior of the
curves are similar to those in Fig.\ 5, but for small $r$ the effect
of phase-space suppression becomes important.

In either case, the conclusion is that for $r$ as large as 2.2, it is
possible to have $\Omega_\pho h^2\la 1$; with our ``central'' choice
of parameters, $[\mu_8^{3}\mu_S^{-4}B]=[\mu_8^{3/2}\mu_S^{-4}C] =1$,
$r$ must be less than $1.8$ in order for $\Omega_\pho h^2\la 1$.
A value of $r$ as small as 1.2 may result in $\Omega_\pho h^2\ga 10^{-2}$;
again with the central choice of parameters the limit is $r\ga1.6$.

Although it apparently is not important for realistic parameters, we
mention a possible special role for the $S^0$, $uds\glu$, the lightest
baryon containing a gluino.  Since the $S^0$ has a non-zero baryon
number, its abundance is {\em not} given by \eqr{eq:EQAB} at low
temperature because of the non-zero baryon number of the Universe.  So
long as the strong interactions are maintaining equilibrium between
nucleons and $S^0$'s, its abundance should be $n_{S^0} \sim
n_N\exp[-(M_{S^0}-m_N)/T]$, where $n_N$ is the nucleon abundance and
$m_N$ is the nucleon mass.  Thus at very low temperature its abundance
will be larger than the $\r0$ abundance, so the co-annihilation and
interconversion processes $\pho S^0\rightarrow K N$ and $\pho N
\rightarrow K S^0$ are a potential sink of $\pho$s which in principle
could help keep the $\pho$ in equilibrium.  However for realistic
cross sections, this does not seem to be important at the relevant
temperatures.  Likewise, although at low enough temperatures there are
more nucleons than pions so that $\Gamma_{\pho N \rightarrow \r0 N}$ is
larger than $\Gamma_{\pho \pi \rightarrow \r0 \pi}$, freeze out has
already occurred before the number density of nucleons begins to
dominate that of pions.

\vspace{36pt}
\centerline{\bf VI. SUMMARY AND CONCLUSIONS}
\vspace{24pt}

We have studied the reactions important for the decoupling and freeze-out
of photinos having mass $m$ less than about 1.5 GeV.  We have found that
it is crucial to include the interactions of the photino with the
$R^0$, the gluon-gluino bound state whose mass $M$ is expected to lie in
the range  1 to 2 GeV.  The $R^0$ has strong interactions and thus
annihilates extremely efficiently and stays in thermal equilibrium to
much lower temperatures.  In this circumstance, photino freeze-out
occurs when the rate of reactions converting photinos to $R^0$'s falls
below the expansion rate of the Universe.  The rate of the $\pho -
\r0$ interconversion interactions which keep photinos in thermal
equilibrium, ($\pho \pi \longleftrightarrow \r0 \pi$) or $\r0$
decay/inverse decay ($\pho \pi \longleftrightarrow \r0$), depends on
the densities of photinos and pions, rather than on the square of the
photino density, as is the case for the self-annihilation process.
For photinos of the relevant mass range ($m \sim 800$ MeV), the pion
abundance is enormous compared to the photino abundance.  Therefore
the photinos stay in equilibrium to much higher values of $x \equiv
m/T$ than they would if self-annihilation were the only operative
process, resulting in a smaller relic density for a given photino mass
and cross section.  We find using the sudden approximation that light
photinos are cosmologically acceptable for a range of $1.2 \la r\equiv
M/m \la 2.2$.  Within this range, if $1.6 \la r \la 2$, the
photinos are an excellent dark matter candidate.  The precise range of
$r$ for which the photino accounts for the cold dark matter may shift
when the sudden approximation is improved and cross sections are
better known.  However the general conclusion is robust:  light
photinos can account for the dark matter of the Universe for a
suitable value of $r$, which is consistent with theoretical
predictions in an attractive class of \SUSY-breaking mechanisms
\re{PHENO}.

Since $\pho$--$\r0$ interconversion governs freeze-out, the usual
relation between $\Omega h^2$ and the relic's annihilation cross
section \re{GW} is not valid.  If inverse decay is the operative
process, then there is no direct prediction for the $\pho$ scattering
cross section on matter.\footnote{Since the short-distance dynamics
entering the matrix element for $R^0 \rightarrow \pho \pi$ is the same
as for the scattering reaction $\pho N \rightarrow R^0 N$, these could
in principle be related.  At this time however we do not have
sufficient control of the hadron physics involved to make a
quantitatively accurate theoretical prediction of the cross sections
from the $\r0$ lifetime.} If $\pho \pi \longleftrightarrow \r0 \pi$ is
the operative process, a quantitative solution of the Boltzmann
equations can be used to infer its cross section.  It will be
significantly smaller, more-or-less by a factor
$n_{\pho}(x_*)/n_{\pi}(x_*)$, than the conventional cross-section used
in planning relic detection experiments.

Direct detection of low-mass relic photinos is more difficult than
detection of high-mass (say $m\sim 50$ GeV) photinos.  In addition to
the low cross section mentioned above, the average energy deposition
is $\avg{E} = m^2M_T\avg{v^2}/(m+M_T)^2$ where $M_T$ is the target
mass. Thus existing and planned experiments using relatively heavy
targets are not well adapted to this search.  On the positive side,
our photino is more likely to have spin-independent couplings to
nucleons than expected in the conventional picture \re{GW}. This is
because in the \SUSY\-breaking mechanism which leads to the light
photino and gluino under discussion here, the off-diagonal terms in
the squark mass-squared matrix can be comparable to the diagonal
terms.\footnote{See Ref.\ \re{PHENO} for allowed ranges of the
parameters determining the squark mass-squared matrix, $\mu,\ \tan
\beta$, and $M_0$.}

Indirect detection via annihilation of gravitationally concentrated
photinos [\ref{INDIRECTA},\ref{INDIRECTB}], for instance trapped in
the Sun, is unlikely.  Because they are low-mass \WIMPS, evaporation
is much more efficient than in the high-mass case, and they do not
concentrate sufficiently.  (And, of course, the cross section is
smaller than conventionally supposed.)

We also note that if the $S^0$ is stable, there will be a relic
abundance of them, with an abundance relative to baryons determined by
$M_{S^0}$ and $x_S \equiv M_{S^0}/T_S$, where $T_S$ is the temperature
of $S^0$ freeze-out.  The $S^0$ mass is expected to be $1.5-2$ GeV, so
let us define $M_{S^0} = 1.5 \mu_{1.5}$ GeV.  Then
\ba
\frac{n_S}{n_B}  &= &\frac{1}{4} \left(\frac{M_{S^0}}{m_N}\right)^{3/2}
                \exp\left[-\frac{M_{S^0}-m_N}{T}\right] \nonumber \\
& = & \frac{1}{4} \left(\frac{1.5\mu_{1.5}\, \GeV}{0.94\, \GeV} \right)^{3/2}
\exp[-(1-0.6/\mu_{1.5})x_S] \ ,
\ea
where the factor $1/4$ accounts for the fact that the $S^0$ is a
spin-0 state and comes in just one flavor, whereas there are 4
spin$\times$flavor degrees of freedom for the baryons.  The $S^0$
self-annihilation cross section should be comparable to that of the
$R^0$, so ignoring the difference between $R^0$ and $S^0$ masses, $x_S
\sim r x_{RR}$, where $x_{RR}$ is the value of $x$ at which
$\Gamma(R^0 R^0 \rightarrow X)/H = 1$.  From Figs.\ 1 and 2 we see
that $r x_{RR} \sim 45$, giving $n_S/n_B \sim 7\times 10^{-9}$ for
$\mu_{1.5} = 1$, and smaller for larger $\mu_{1.5}$.  Since the
$S^0$'s are strongly interacting, even this small an abundance may be
detectable.  They will be more gravitationally concentrated than
standard \WIMPS\ of comparable mass because they dissipate energy
through their strong interactions, although they do not form atoms or
bind to nuclei\footnote{If they were stable and could bind to nuclei,
they would have been detected in rare isotope searches\re{EXPTS}, so
that possibility is excluded.}.

What, then, is the strategy for testing the proposal that photinos
with mass less than or about 1 GeV constitute the cold dark matter of
the Universe?  Of course if an $R^0$ in the 1 to 2 GeV range could be
excluded by laboratory searches, our suggestion for the dark matter
would be also excluded.  Assuming though that these particles are
discovered, knowledge of experimentally accessible properties of the
photino and $R^0$ (in particular, their masses, the $R^0$ lifetime,
and the cross section for $R^0 N \rightarrow \pho N$) coupled with
detailed numerical analysis of the freeze-out process, will allow a
much more accurate prediction of the relic abundance than has been
possible here.  Since the relic density is exponentially dependent on
$r$, which will one day be well measured, an accurate quantitative
test of this idea will eventually be possible.

In the meantime, theoretical work can elucidate the viability of this
proposal.  In the \SUSY-breaking mechanism relevant to this scenario
the parameters $\mu,\ \tan \beta$, and $M_0$, which determine the
photino and gluino masses, are highly constrained \re{PHENO}.
Relatively soon even more accurate predictions for the photino and
gluino masses will be possible, narrowing the possible range of
photino masses corresponding to any allowed gluino mass.  With use of
lattice gauge theory, it should be possible to compute the $R^0$ mass
range corresponding to a given gluino mass, and thus to determine a
spectrum of possible $r$ values.  Lattice gauge calculations can also
in principle determine the matrix element for $R^0 \rightarrow \pho
\pi$, given the squark masses (which are fixed by the same unknown
parameters determining $m_{\pho}$ and $m_{\glu}$, at least with
minimal \SUSY\ breaking), and determine the masses of the other
$R$-hadrons, which would help in estimating the cross sections.  For
instance knowledge of the mass of the $R_{\pi}$ would allow one to
better model $\sigma_{\r0\pi}$.  With more accurately fixed inputs, a
full numerical solution of the coupled Boltzmann equations would be
justified.

Therefore, the most important next steps are:
\begin{enumerate}
\item Look hard for $R$-hadrons and other new particles predicted by
this scenario.  Planned kaon experiments may be able to establish
evidence for the $R^0$, and possibly measure the its lifetime and
mass, as well as the mass of the photino \re{PHENO}.
\item Do a better job fixing the parameters of the underlying theory,
as well as calculating the photino mass  produced through radiative
corrections.
\item Use lattice gauge theory to calculate the $R^0$ mass and
check other predictions of this scenario such as the origin of the
$\eta'$ mass \re{PHENO}.
\item A more complete treatment  modeling the
photino freeze-out is necessary \re{CFK}.  An immediate question to
address is the quality of the sudden approximation used here.  When
interconversion is the dominant process, the equation governing the
evolution of the photino density has a somewhat different form than in
the self-annihilation case, for which the quality of the sudden
approximation is well-established.
\item  Obtain detailed predictions for the low energy $\pho$-nucleus
cross sections expected in this scenario, and find effective detection
techniques for light photinos.
\end{enumerate}

At the very least we have shown that until the value of $r$ is
demonstrated to be larger than about 2.2, light photinos are
cosmologically acceptable.  At best, we have described the scenario for
the production and survival of the dark matter of the Universe.

While there is no shortage of candidates for relic dark matter
particle species, this proposal extends the idea that photinos may be
the dark matter to a previously excluded mass range by incorporating
new reactions that determine the photino relic abundance.   If this
scenario is correct, direct  and indirect detection of dark matter
might be even more difficult than anticipated.  However the scenario
requires the existence of low-mass hadrons, which can be produced and
detected at  accelerators of moderate energy.  Thus particle physics
experiments will either disprove this scenario, or make light photinos
the leading candidate for dark matter.

\vspace{36pt}

\centerline{\bf ACKNOWLEDGMENTS}

GRF was supported in part by the NSF (NSF--PHY--91--21039).
EWK is supported by the DOE and NASA under Grant NAG5--2788.

\frenchspacing
\def\prpts#1#2#3{Phys. Reports {\bf #1}, #2 (#3)}
\def\prl#1#2#3{Phys. Rev. Lett. {\bf #1}, #2 (#3)}
\def\prd#1#2#3{Phys. Rev. D {\bf #1}, #2 (#3)}
\def\plb#1#2#3{Phys. Lett. {\bf #1B}, #2 (#3)}
\def\npb#1#2#3{Nucl. Phys. {\bf B#1}, #2 (#3)}
\def\apj#1#2#3{Astrophys. J. {\bf #1}, #2 (#3)}
\def\apjl#1#2#3{Astrophys. J. Lett. {\bf #1}, #2 (#3)}
\begin{picture}(400,50)(0,0)
\put (50,0){\line(350,0){300}}
\end{picture}

\vspace{0.25in}

\def\labelenumi{[\theenumi]}

\begin{enumerate}

\item \label{EXPTS} G. R. Farrar, \prd{51}{3904}{1995}.

\item\label{PHENO} G. R. Farrar,  Technical Report RU-95-17
(hep-ph/9504295),  Rutgers Univ., 1995.

\item \label{BKN}  T. Banks, D. B. Kaplan, A. E.  Nelson, \prd{49}{779}{1994}.

\item \label{FM} G. R. Farrar and A. Masiero, Technical Report RU-94-38
(hep-ph/9410401),  Rutgers Univ., 1994.

\item \label{pierce_papa}
D. Pierce and A. Papadopoulos, \npb{430}{278}{1994}.

\item \label{bgm}
R. Barbieri, L. Girardello, and A. Masiero,  \plb{127}{429}{1983}.

\item \label{bm}
R. Barbieri and L. Maiani,  \npb{243}{429}{1984}.

\item \label{G}  H. Goldberg, \prl{50}{1419}{1983}.

\item \label{LSP} J. Ellis, J. S. Hagelin, D. V. Nanopolous, K. Olive, and
M. Srednicki,  \npb{238}{453}{1984};
G. B. Gelmini, P. Gondolo, and E. Roulet,  \npb{351}{623}{1991};
K. Griest, \prd{38}{2357}{1988};
L. Rozkowski, \plb{262}{59}{1991}.

\item \label{GF:84} G. R. Farrar, \prl{53}{1029}{1984}.

\item \label{KT} E. W. Kolb and  M. S. Turner, {\em The Early Universe},
(Addison-Wesley, Redwood City, Ca., 1990).

\item \label{KR}  E. W. Kolb and S. Raby,  \prd{27}{2990}{1983}.

\item \label{GG} P. Gondolo and G. Gelmini, \npb{360}{145}{1991}.

\item \label{INDIRECTA} J. Silk, K. Olive, and M. Srednicki,
\prl{53}{624}{1985};
T. K. Gaisser, G. Steigman, and S. Tilav, \prd{34}{2206}{1986};
J. S. Hagelin, K. W. Ng, and K. A. Olive,  \plb{180}{375}{1987};
M. Srednicki, K. A. Olive, and J. Silk, \npb{279}{804}{1987}.

\item\label{KKC}  G. L. Kane and I. Kani, \npb{277}{525}{1986};
B. A. Campbell {\em et al.}, \plb{173}{270}{1986}.

\item\label{HK} H. Haber and G. Kane, \prpts{117}{75}{1985}.

\item \label{CFK} D.  J. Chung,  G. R. Farrar, and E. W. Kolb, in preparation.

\item \label{GW} M. W. Goodman and E. Witten, \prd{31}{3059}{1985};
I. Wasserman, \prd{33}{2071}{1986}.

\item \label{INDIRECTB} M. L. Krauss,  M. Srednicki, and F. Wilczek,
        \prd{33}{2079}{1986};
K. Freese, \plb{167}{295}{1986}.

\end{enumerate}

\end{document}